\documentclass[9pt,conference]{IEEEtran}

\usepackage{waspaa25} 

\usepackage{bm} 

\usepackage{algorithm}
\usepackage{algpseudocode}
\algnewcommand{\Inputs}[1]{%
  \State \textbf{Inputs:}
  \Statex \hspace*{\algorithmicindent}\parbox[t]{.8\linewidth}{\raggedright #1}
}
\algnewcommand{\Initialize}[1]{%
  \State \textbf{Initialize:}
  \Statex \hspace*{\algorithmicindent}\parbox[t]{.8\linewidth}{\raggedright #1}
}

\newcommand{\h}{\mathbf{h}}
\newcommand{\y}{\mathbf{y}}
\newcommand{\x}{\mathbf{x}}
\newcommand{\n}{\mathbf{n}}
\newcommand{\refsub}{\mathrm{ref}}
\newcommand{\occsub}{\mathrm{o}}
\newcommand{\unoccsub}{\text{\o}}
\newcommand{\nosesub}{1}
\newcommand{\Ry}{\mathbf{R}_y}
\newcommand{\Rx}{\mathbf{R}_x}
\newcommand{\Rn}{\mathbf{R}_n}
\newcommand{\Ryest}{\hat{\mathbf{R}}_y}

\newcommand{\Rnest}{\hat{\mathbf{R}}_n}

\newcommand{\Rxocc}{\mathbf{R}_{x,\occsub}}
\newcommand{\Rnocc}{\mathbf{R}_{n,\occsub}}

\newcommand{\Rxunocc}{\mathbf{R}_{x,\unoccsub}}

\newcommand{\phix}{\phi_x}

\newcommand{\Bmat}{\mathbf{B}}
\newcommand{\Gmat}{\mathbf{G}}

\newcommand{\w}{\mathbf{w}}
\newcommand{\hocc}{\h_\occsub}
\newcommand{\hunocc}{\h_\unoccsub}
\newcommand{\hest}{\hat{\h}}

\newcommand{\Ryoccest}{\hat{\mathbf{R}}_{y,\occsub}}

\newcommand{\Rnoccest}{\hat{\mathbf{R}}_{n,\occsub}}
\newcommand{\Ryunoccest}{\hat{\mathbf{R}}_{y,\unoccsub}}

\newcommand{\Rnunoccest}{\hat{\mathbf{R}}_{n,\unoccsub}}
\newcommand{\Rnnuest}{\hat{\mathbf{R}}_{n,\nu}}
\newcommand{\Rynuest}{\hat{\mathbf{R}}_{y,\nu}}
\newcommand{\inv}{^{-1}}

\newcommand{\OVD}{{\rm OVD}}

\newcommand{\Rntilde}{\tilde{\mathbf{R}}_{n,\unoccsub}}

\usepackage[subtle]{savetrees}

\title{Microphone Occlusion Mitigation for Own-Voice Enhancement in Head-Worn Microphone Arrays Using Switching-Adaptive Beamforming}

\name{Wiebke Middelberg$^{1,2}$\sthanks{This work was done during an internship at Meta Reality Labs.},
      Jung-Suk Lee$^{1}$,
      Saeed Bagheri Sereshki$^{1}$,
      Ali Aroudi$^{1}$,
      Vladimir Tourbabin$^{1}$,
      Daniel D. E. Wong$^{1}$}
\address{$^{1}$Meta Reality Labs, Redmond, WA, USA\;
$^{2}$Carl von Ossietzky Universit\"at Oldenburg,
Oldenburg, Germany
}

\begin{document}

\maketitle

\begin{abstract}
Enhancing the user's own-voice for head-worn microphone arrays is an important task in noisy environments to allow for easier speech communication and user-device interaction. However, a rarely addressed challenge is the change of the microphones' transfer functions when one or more of the microphones gets occluded by skin, clothes or hair.
The underlying problem for beamforming-based speech enhancement is the (potentially rapidly) changing transfer functions of both the own-voice and the noise component that have to be accounted for to achieve optimal performance.
In this paper, we address the problem of an occluded microphone in a head-worn microphone array. We investigate three alternative mitigation approaches by means of (i) conventional adaptive beamforming, (ii) switching between a-priori estimates of the beamformer coefficients for the occluded and unoccluded state, and (iii) a hybrid approach using a switching-adaptive beamformer.
In an evaluation with real-world recordings and simulated occlusion, we demonstrate the advantages of the different approaches in terms of noise reduction, own-voice distortion and robustness against voice activity detection errors.

\end{abstract}

\section{Introduction}
Head-worn microphone arrays like hearing aids, most modern headphones, and smart or virtual reality glasses can be effective at capturing the user's own-voice due to the vicinity to the source. This is beneficial for speech communication, e.g., for telephony or human-machine interaction using voice commands \cite{levin2016nearfield,Hoang2022OwonVoice,ohlenbusch2024Occlusion}. Despite the fact that even in noisy environments the user's own-voice might already be captured at a relatively high signal-to-noise ratio (SNR), noise reduction algorithms, such as beamformers steering towards the user's mouth, can further help to improve speech intelligibility \cite{benesty2008springer,Doclo2015,levin2016nearfield,Gannot2017,Hoang2022OwonVoice,ohlenbusch2024Occlusion}.

A common problem for body-worn microphone arrays is susceptibility to user movement, i.e., inducing changes in the microphones' relative transfer functions due to deformation of the array \cite{Corey2019Motion} or quick movements relative to the acoustic scene \cite{Casebeer2021NiceBeam}. Especially fixed spatial filters, typically relying on known microphone array geometries and microphone transfer functions, can experience a degradation of performance due to changes in the array characteristics \cite{VanVeen1988a,Ehrenberg2010,Tesch2023JNF,mannanova2024meta}, e.g., due to deformation. 
A re-calibration procedure for changing transfer functions used in a generalized sidelobe canceler was proposed in \cite{Oak2005calibration}.
Another problem of body- or head-worn microphone arrays, also causing potentially rapid changes in the transfer functions, is the risk of certain microphones being occluded by skin, hair or clothing, often resulting in a muffled, i.e., low-pass filtered, sound, which can affect the performance of speech enhancement system. The problem of occluded microphones for array processing was considered in \cite{Bando2016MicOcc} where a deformable array with partial occlusion was addressed.
In the context of dereverberation, the problem of rapidly changing transfer functions/filter vectors was considered in \cite{ikeshita2021onlineSwitchingWPE}, where a switching version of the adaptive weighted prediction error algorithms was proposed. 
Switching beamformers with adaptive noise covariance matrices per filter were proposed in \cite{Yamaoka2019switching} to address the problem of interferer reduction in underdetermined situations. Examples for dictionary-based approaches for rapid dynamics are e.g. wind noise suppression \cite{Tammen2022dictionary} or automatic speech recognition with varying source directions \cite{ito2017probabilistic}.

In this paper, we investigate the problem of own-voice enhancement for a head-worn microphone array, where a particular microphone is prone to being sporadically and dynamically occluded. 
As we regard the detection of the occlusion as a separate problem \cite{madhu2011SensorAnomaly,Gaal2019MicBlockage}, we base our work on the assumption that a reliable occlusion detector is available. We propose the following processing strategies to mitigate the effect of the occlusion: (i) a standard implementation of an adaptive beamformer, (ii) a switching mechanism between a-priori estimates of the filter vectors for the occluded and unoccluded state, i.e., similar to a dictionary-based approach, and (iii) a hybrid switching-adaptive beamformer which adapts two sets of covariance matrices depending on the occlusion state.

The proposed algorithms are evaluated using real-world recordings of own-voice in noise with dynamically changing occlusions. 
The evaluation is performed on multiple speech and noise samples with different dynamics in the occlusion pattern and multiple SNRs.
We demonstrate the advantages of the different processing strategies in terms of noise reduction, own-voice distortion and robustness against a non-optimal voice activity detection (VAD). The results show the potential of the proposed switching-adaptive beamformer, which exhibits lower own-voice distortions for highly dynamic changes in the occlusion state than a conventional adaptive beamformer, while clearly outperforming the purely switching beamformer in terms of SNR improvement if a good VAD is available.

\section{Signal Model and Problem Formulation}

We consider a microphone configuration of a head-worn microphone array with $M$ microphones, which capture the user's own-voice (considered as the target signal) and farfield noise. The noisy microphone signal in the $m$-th microphone in the discrete Fourier transform (DFT) domain can be written as
\begin{equation}\label{eq:SigM}
    Y_m(k,t) = X_m(k,t) + N_m(k,t)\, , \quad m \in \{1,\dots,M\}\, ,
\end{equation}
where $k$ and $t$ denote the frequency bin and time frame index respectively, and $X_m(k,t)$ and $N_m(k,t)$ are the speech and farfield noise captured by the $m$-th microphone, respectively. As all frequency bins and time frames are assumed to be independent and are processed as such, we will neglect $k$ and $t$ in the remainder of the paper wherever possible. The signal model in \eqref{eq:SigM} can be written in terms of the $M$-dimensional signal vector $\y= [Y_1, Y_2,\dots,Y_M]^T$ containing all microphone signals, where $\{\cdot\}^T$ denotes the transpose operator, i.e.,
\begin{equation}\label{eq:SigVec}
    \y = \x + \n\, ,
\end{equation}
where the speech and noise vector $\x$ and $\n$ are defined similarly to $\y$.
For the speech component, a multiplicative transfer function is assumed \cite{Avargel2007multiplicative}, allowing to write the speech vector as
\begin{equation}\label{eq:RTF}
    \x = \h X_\refsub\, ,
\end{equation}
where $X_\refsub$ denotes the speech component in the reference microphone, and the RTF vector $\h$ contains the ratios of acoustic transfer functions (ATFs) to all microphones ($A_1,\dots,A_M$) relative to a reference microphone, such that the entry of the reference microphone is 1 by definition, i.e.,
\begin{equation}\label{eq:DefRTF}
\begin{split}
    \h &= [A_1/A_\refsub,A_2/A_\refsub,\dots,1,\dots,A_M/A_\refsub]^T\, .
\end{split}
\end{equation} 

Assuming statistical independence of the speech and noise component, the noisy covariance matrix can be written as 
\begin{equation}\label{eq:Ry}
    \begin{split}
        \Ry &= \mathcal{E}\{\y\y^H \}= \Rx + \Rn\, ,
    \end{split}
\end{equation}
where $\mathcal{E}\{\cdot\}$ denotes the expectation operator, $\{\cdot\}^H$ denotes the Hermitian transpose operator, and $\Rx$ and $\Rn$ are the speech and noise covariance matrix, respectively.
Using \eqref{eq:RTF}, $\Rx$ can be written as a rank-1 matrix spanned by the RTF vector i.e.,
\begin{equation}\label{eq:Rx}
    \Rx = \phix \h\h^H\, ,
\end{equation}
which is scaled by the speech power spectral density in the reference microphone $\phix = \mathcal{E}\{|X_\refsub|^2\}$.
For the noise component, we assume the covariance matrix $\Rn$ to be full-rank.

\begin{figure}[t]
\centerline{\includegraphics[width = \linewidth]{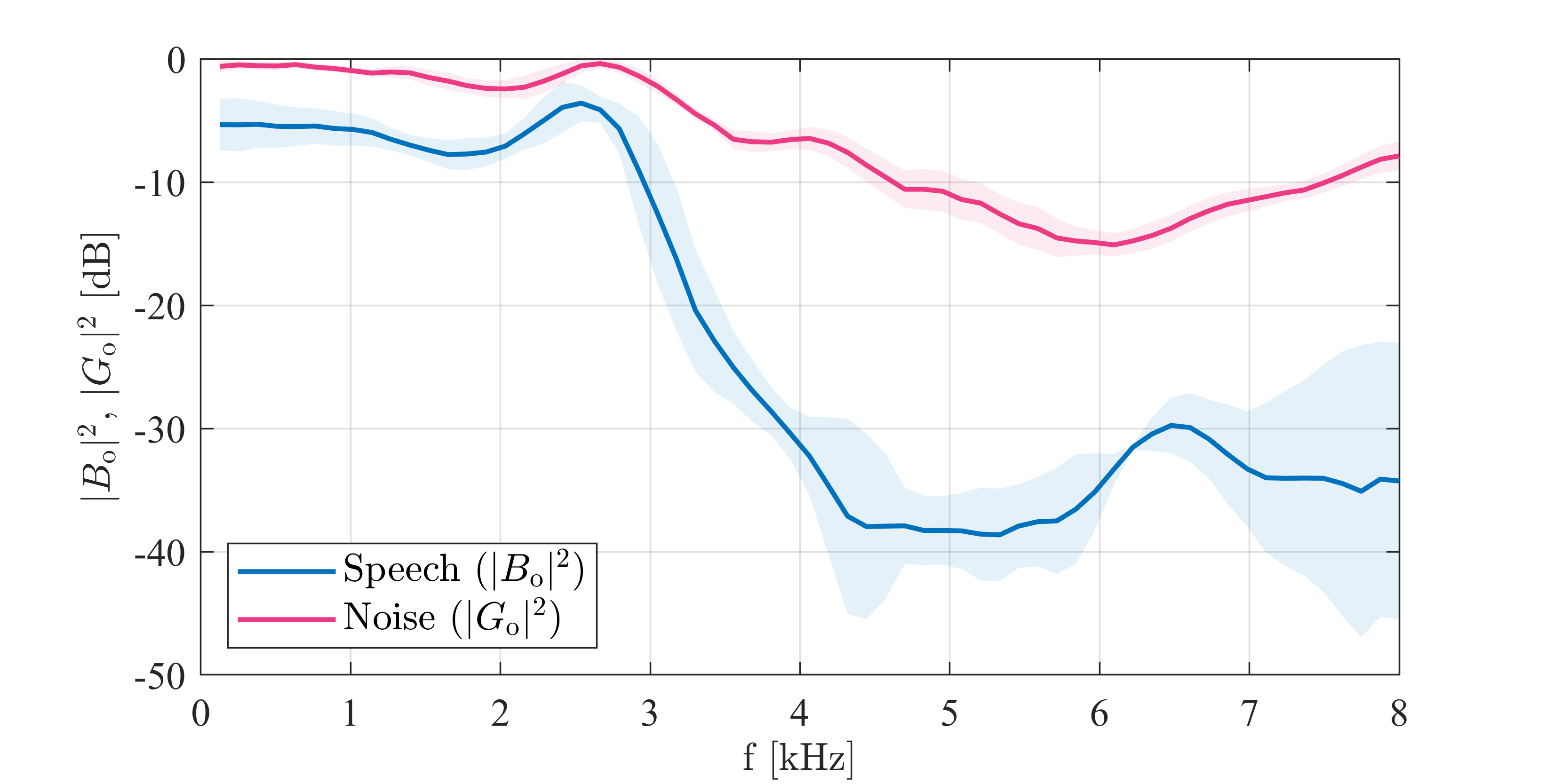}}
\caption{Nearfield (own-voice) and farfield (noise) transfer function between occluded and unoccluded state averaged over multiple users and sound fields (solid lines) and their standard deviations (shaded areas).}
\label{fig:AllTFs}
\end{figure}

To include the potential occlusion of a certain microphone into the signal model, we define the first microphone to be the potentially occluded one, without loss of generality. Furthermore, we define the two ATFs $A_\nosesub=A_\unoccsub$ for the unoccluded default state in \eqref{eq:DefRTF} and $A_\occsub$ for the occluded state (and similar for the unoccluded and occluded RTFs $H_\unoccsub$ and $H_\occsub$), respectively.
The relation between these two states of the first microphone can be defined as 
\begin{equation}\label{eq:Xocc}
    X_{\nosesub,\occsub} = B_\occsub X_{\nosesub,\unoccsub}\, , \quad N_{\nosesub,\occsub} = G_\occsub N_{\nosesub,\unoccsub}\, ,
\end{equation}
where $X_{\nosesub,\occsub}$, $X_{\nosesub,\unoccsub}$, $N_{\nosesub,\occsub}$ and $N_{\nosesub,\unoccsub}$ are the occluded and unoccluded speech and noise component in the first microphone, respectively. $B_\occsub$ and $G_\occsub$ are the occlusion transfer functions, i.e., the RTFs between the occluded and unoccluded state of the first microphone, where the occlusion transfer function for the speech component can be written as $B_\occsub = A_\occsub/A_\unoccsub = H_\occsub/H_\unoccsub$, for which it should be noted that the definition in terms of the RTFs $H_\occsub$ and $H_\unoccsub$ only holds if the first microphone is not selected as the reference microphone.
Also note that even though we cannot write the noise component $\n$ in terms of ATFs (or an RTF vector), we can still define the transfer function $G_\occsub$ between the unoccluded and occluded state of the first microphone. 
Using \eqref{eq:Xocc}, the occluded RTF vector $\hocc$ can be written as
\begin{equation}\label{eq:RTFtrafo}
    \hocc = \Bmat \hunocc\, ,
\end{equation}
where $\hunocc$ is the unoccluded RTF vector and the transformation matrix $\Bmat$ is defined as
\begin{equation}\label{eq:Bmat}
    \Bmat = {\rm diag}([B_\occsub,\mathbf{1}_{M-1}^T])\; ,
\end{equation}
where ${\rm diag}(\cdot)$ creates a diagonal matrix out of a vector and $\mathbf{1}_{M-1}$ is the $(M-1)$-dimensional vector of ones.
Using \eqref{eq:Bmat}, the occluded speech covariance matrix is given by
\begin{equation}\label{eq:Rxocc}
    \Rxocc = \Bmat\Rxunocc\Bmat^H = \phix \Bmat\hunocc\hunocc^H\Bmat^H\, .
\end{equation}
The transformation matrix $\Gmat$ and the occluded noise covariance matrix $\Rnocc$ are defined similarly to \eqref{eq:Bmat} and \eqref{eq:Rxocc}.
Generally, the two occlusion transfer functions $B_\occsub$ and $G_\occsub$ for the speech and noise component do not have to be the same.
\Cref{fig:AllTFs} depicts the power of the occluded speech and noise transfer function respectively for our specific case\footnote{The RTFs $B_\occsub$ and $G_\occsub$ between the occluded and unoccluded state were obtained from recordings of clean speech and noise on multiple users (data from 16 users contributed to the extraction of the transfer functions). For the speech component, the occluded RTF was obtained from the occluded and unoccluded RTF vector, while for the noise component, only a relative gain was extracted from the occluded and unoccluded noise covariance matrix.}, clearly showing different transfer characteristics for the speech and the noise component.
Such differences in the effect of occlusion on different signal components can, for example, be caused by the directionality of the occluded microphone.

The output signal of the multi-channel filter is defined as $Z = \w^H\y$ with the $M$-dimensional filter vector $\w$ for which we will use the minimum-variance distortionless response (MVDR) beamformer\cite{Doclo2015,Gannot2017,VanVeen1988a}, i.e.,
\begin{equation}\label{eq:MVDR}
    \w_\nu = \frac{\Rnnuest\inv\hest_\nu}{\hest_\nu^H\Rnnuest\inv\hest_\nu}\, , \quad \nu \in \{\unoccsub,\occsub\}\, ,
\end{equation}
where a hat denotes an estimate of a quantity and the beamformer depends on the occlusion state $\nu$.

\section{Occlusion Mitigating Processing}\label{sec:Proc}

In this section, we discuss possible approaches on how to mitigate the influence of the occlusion on an MVDR beamformer. In \cref{subsec:AdaptMVDR}, we present the state-of-the-art processing for dynamic acoustic scenarios, i.e., adaptive covariance matrix and RTF vector estimation. In \cref{subsec:SwitchingMVDR}, we introduce a switching mechanism for the occluded and unoccluded state, making use of a-priori known transfer functions. In \cref{subsec:HybridProcessing}, we propose a combination that adapts different sets of covariance matrices for the occluded and unoccluded state, respectively.

\subsection{Adaptive Beamforming}\label{subsec:AdaptMVDR}

In many speech enhancement algorithms for dynamic scenes, the noisy and noise covariance matrix are estimated by means of recursive smoothing \cite{Ephraim1984,GoesslingICASSP2019,Braun2016dynamicKalman,donley2021adaptive}, i.e.,
\begin{equation}\label{eq:RySmooth}
\Ryest(t) =
\begin{cases}
     \alpha_y \Ryest(t-1) + (1-\alpha_y) \y(t)\y(t)^H\, ,\hspace{-0.25cm} &\text{ if VAD($t$) = 1}\\
    \Ryest(t-1)\, , &\text{ if VAD($t$) = 0},
\end{cases}
\end{equation}
\begin{equation}\label{eq:RnSmooth}
    \Rnest(t) = 
    \begin{cases}
    \alpha_n \Rnest(t-1) + (1-\alpha_n) \y(t)\y(t)^H\, ,\hspace{-0.3cm} &\text{ if VAD($t$) = 0}\\
     \Rnest(t-1)\, ,\; &\text{ if VAD($t$) = 1},
    \end{cases}
\end{equation}
with the smoothing constants $\alpha_y$ and $\alpha_n$, where the updates depend on the (binary) voice activity detection (VAD).

For the adaptive estimation of the RTF vector $\hest$, a power method implementation \cite{golub2013matrix} of the generalized eigenvalue decomposition (GEVD)-based RTF estimation \cite{Markovich2009,Krueger2010GSClike,Serizel2014,Cohen2022GEVD} is employed. The RTF vector is estimated as the (normalized and de-whitened) principal eigenvector of the matrix pencil ($\Rnest,\Ryest$).

Assuming that a fast adaptation is sufficient to capture the potentially highly time-varying dynamics of occlusion, the adaptive beamforming techniques described above should be able to account for occlusion effects after a short adaptation period. However, it should also be noted that only the noisy or the noise covariance matrix can be updated at a time, meaning that if occlusion occurs e.g. while speech is active, the noise covariance matrix cannot adapt to the occlusion. This further implies that frequent and rapid changes might not fully be captured and hence the beamformer might be suboptimal in a sense that the estimated covariance matrices do not reflect the true signal characteristics. 
Furthermore, note that adaptive processing as described above relies on a (good) VAD, while not depending on an occlusion detection.

\subsection{Switching Beamforming}\label{subsec:SwitchingMVDR}

Another approach to handling occlusion makes use of the a-priori knowledge of both occluded and unoccluded RTF vectors and occluded and unoccluded noise covariance matrices, which are switched, as indicated in \eqref{eq:MVDR}, depending on the detected occlusion state. 

Hence, we assume that the occluded and unoccluded relative transfer functions, i.e., an a-priori estimate, denoted by a tilde, of the (unoccluded) RTF vector $\tilde{\h}_\unoccsub$ (cf. \eqref{eq:RTF}) and the occluded transfer functions for speech and noise $\tilde{B}_\occsub$ and $\tilde{G}_\occsub$ in \eqref{eq:Xocc}, respectively, are available. Furthermore, the unoccluded noise covariance matrix is modeled as diffuse as the a-priori estimate $\Rntilde$. Using the transformation in \eqref{eq:Rxocc} allows for computing the occluded RTF vector and noise covariance matrix. These assumptions are realistic to make for a known, fixed array where the user's own-voice is the signal of interest, which comes from an well known position and where prior measurements of transfer functions can be performed.
Effectively, the used beamformer $\w$ is either of the two fixed filter vectors $\tilde{\w}_\unoccsub$ or $\tilde{\w}_\occsub$ (cf. \eqref{eq:MVDR}, computed based on the a-priori estimates), depending on the binary occlusion detection (OD), i.e.,
\begin{equation}\label{eq:SwMVDR}
    \w(t) = 
    \begin{cases}
    \tilde{\w}_\unoccsub\, ,\; &\text{ if OD($t$) = 0}\\
    \tilde{\w}_\occsub\, ,\; &\text{ if OD($t$) = 1}\, .
    \end{cases}
\end{equation}
As this processing entirely relies on a-priori knowledge and does not adapt to the acoustic scenario (only to occlusion), the used estimates might not fit the data ideally, and might hence not lead to optimal performance. This means that e.g. user variability, model mismatches or changes in the acoustic scene cannot be accounted for. However, this processing also comes with a certain robustness against VAD errors, as it does not rely on a VAD.

\subsection{Hybrid Switching-Adaptive Beamforming}\label{subsec:HybridProcessing}

To overcome the limitations of the two approaches in \cref{subsec:AdaptMVDR} and \cref{subsec:SwitchingMVDR}, we propose a combination, depending on both the VAD and the OD.
Instead of adapting the covariance matrices independent of the occlusion state as in \eqref{eq:RySmooth} and \eqref{eq:RnSmooth}, we propose to adapt two different sets of covariance matrices depending on the occlusion state. The update rule in \eqref{eq:RySmooth} for the noisy covariance matrix can hence be formulated as
\begin{equation}\label{eq:}
    \Rynuest(t) = \begin{cases}
     \alpha_y \Rynuest(t_\nu) + (1-\alpha_y) \y(t)\y(t)^H\, ,\hspace{-0.3cm} &\text{ if VAD($t$) = 1}\\
    \Rynuest(t_\nu)\, , &\text{ if VAD($t$) = 0},
\end{cases}
\end{equation}
where $\nu$ corresponds to the currently detected occlusion state and $t_\nu$ is the frame where the respective occlusion state was detected last. A similar update rule can be formulated for the noise covariance matrix $\Rnnuest(t)$ as in \eqref{eq:RnSmooth}.
The processing is summarized in \Cref{alg:SwAdapt}. 

It should be noted that a practical occlusion detection and the second set of covariance matrices increase the computational complexity and memory consumption compared to the adaptive beamformer which is a caveat for resource-constrained on-device applications.

\begin{algorithm}
\caption{Switching-adaptive covariance estimation}\label{alg:SwAdapt}
\begin{algorithmic}
\Inputs{$\tilde{\h}_\unoccsub$, $\Rntilde$, $\tilde{\Bmat}$, $\tilde{\Gmat}$}
\Initialize{\strut 
$t_\unoccsub \gets 0$, $t_\occsub \gets 0$\\
$\Ryunoccest(0) \gets \tilde{\h}_\unoccsub \tilde{\h}_\unoccsub^H$,\quad $\Ryoccest(0) \gets \tilde{\Bmat}\Ryunoccest(0)\tilde{\Bmat}^H$\\
$\Rnunoccest(0) \gets \Rntilde$,\quad $\Rnoccest(0) \gets \tilde{\Gmat}\Rnunoccest(0)\tilde{\Gmat}^H$}
\vspace{0.2cm}
\For{$t = 1$ to $T$}

\State$\nu\! \gets\! \begin{cases}
     \unoccsub , &\hspace{-0.35cm}\text{ if OD($t$) = 0}\\
    \occsub\, , &\hspace{-0.35cm}\text{ if OD($t$) = 1}
\end{cases}$

    \State$\Rynuest(t)\! \gets\! \begin{cases}
     \alpha_y \Rynuest(t_\nu)\! +\! (1\!-\!\alpha_y) \y(t)\y(t)^H , &\hspace{-0.35cm}\text{ if VAD($t$) = 1}\\
    \Rynuest(t_\nu)\, , &\hspace{-0.35cm}\text{ if VAD($t$) = 0}
\end{cases}$
\State $\Rnnuest(t)\! \gets\! \begin{cases}
     \alpha_y \Rnnuest(t_\nu)\! +\! (1\!-\!\alpha_y) \y(t)\y(t)^H,\! &\hspace{-0.35cm}\text{ if VAD($t$) = 0}\\
    \Rnnuest(t_\nu)\, , &\hspace{-0.35cm}\text{ if VAD($t$) = 1}
\end{cases}$
    \State $\hest_\nu \gets \mathcal{P}\{\Rnnuest(t)^{-1}\Rynuest(t)\}$ \Comment{GEVD-based RTF est.}
    \State $\w(t) \gets \frac{\Rnnuest(t)\inv\hest_\nu}{\hest_\nu^H\Rnnuest(t)\inv\hest_\nu}$
    \State $t_\nu \gets t$ 
\EndFor
\end{algorithmic}
\end{algorithm}

\section{Evaluation}

\begin{figure*}[t]
\centerline{\includegraphics[width = \textwidth, trim={2cm 0.4cm 2cm 0.3cm},clip]{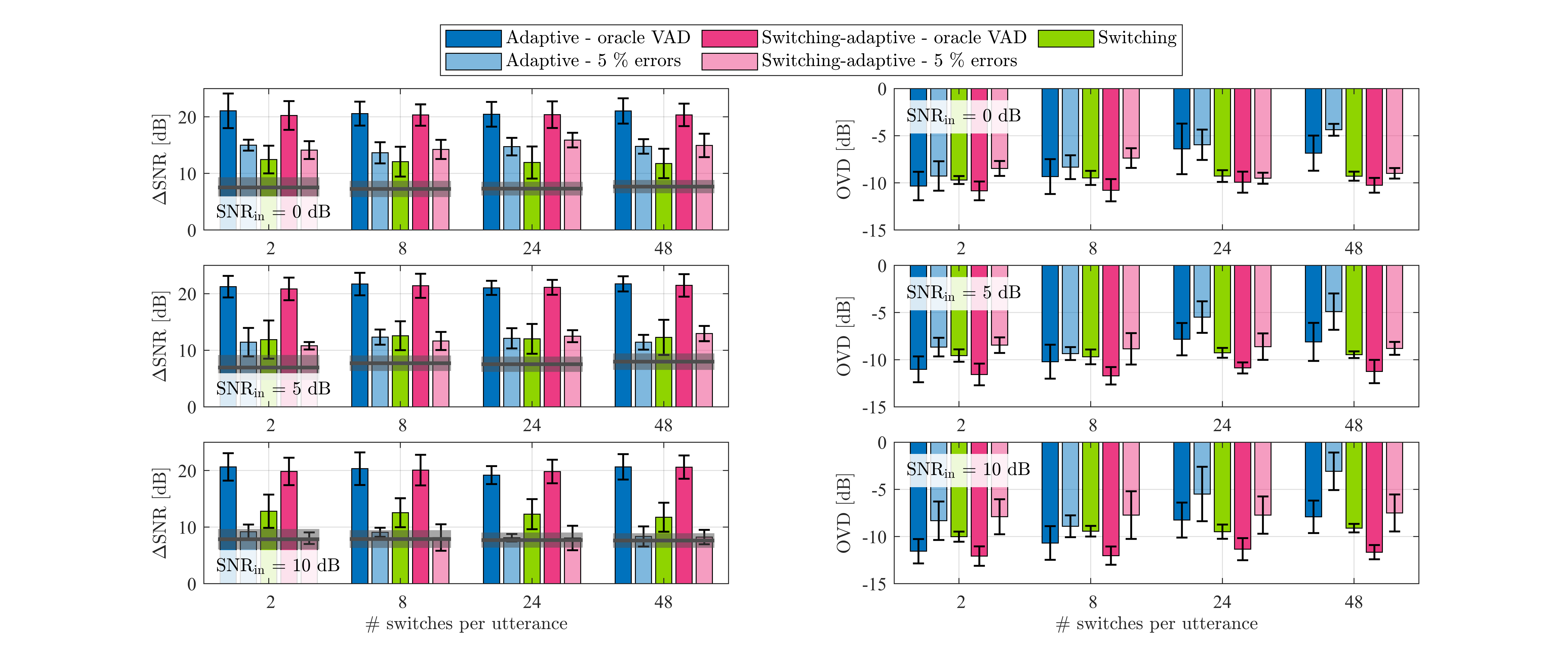}}
\caption{Results for evaluation with two different VADs (oracle and with 5\% false negatives) for different input SNRs and numbers of switches in the occlusion state per utterance. Left panels: SNR improvement with occluded microphone as reference line. Right panels: own-voice distortions.}
\label{fig:ResultsEval}
\end{figure*}

\subsection{Simulation Framework and Conditions}
The evaluation of the proposed algorithms was conducted using real-world recordings of speech and diffuse-like noise, recorded separately in an acoustically treated room at a sampling rate of 16 kHz.
A total of six noisy signals were used for the evaluation, consisting of three different speech signals and two different noise signals, each with a duration of about 13 s.
A five-channel head-mounted array in the form factor of glasses was used, with one microphone on the nose pad affected by occlusion. Speech and noise were mixed at input SNRs of 0, 5, and 10 dB in the reference microphones close to the user's ears. To simulate authentic occlusion patterns, the occlusion transfer functions (see \cref{fig:AllTFs}) were imposed on the unoccluded microphone signals before mixing.
Random occlusion patterns with varying numbers of switches (2, 8, 24, and 48) per utterance were generated to investigate the effect of different dynamics of occlusion. 

The processing was performed in the DFT domain using a weighted overlap add framework with an effective frame length of 16 ms and an overlap of 75\%, analysis and synthesis used custom windows to reduce leakage effects. The smoothing constants $\alpha_y$ and $\alpha_n$ for the noisy and noise covariance matrix corresponded to forgetting times of 0.3 s and 0.5 s, respectively.
An oracle occlusion detection (as the detection of occlusion is beyond the scope of this paper) and VAD were used, with the latter also tested with 5\% artificially induced false negatives to evaluate robustness against VAD errors.

The performance of the algorithms was evaluated in terms of binaural SNR improvement and own-voice distortion (OVD) relative to the reference channels. 
For the OVD, we employ the negative scale-invariant signal-to-distortion ratio \cite{SISDR}, i.e.,
\begin{equation}\label{eq:SIRSDR}
    \OVD = -20\;  \log_{10} \left(\frac{||c x_\refsub||_2}{||c x_\refsub - x_{\rm out}||_2}\right)\, ,
\end{equation}
with $c = x_{\rm out}^Tx_\refsub/||x_\refsub||_2^2$ and the time domain sequences $x_\refsub$ and $x_{\rm out}$ as the reference input signal and the filtered output signal. 
Both objective measures are computed on the time domain signals during speech activity, and are averaged over the left and right side.

\subsection{Results}

The evaluation results are shown in \cref{fig:ResultsEval}, where the panels on the left depict the SNR improvement (where higher is better), and the panels on the right depict the OVDs (where lower is better).
The performance is plotted over the numbers of switches in the occlusion state per utterance. The different bars show the mean results over all utterances for the different algorithms (adaptive, switching and switching-adaptive), where the solid colors represent the results for an oracle VAD, while the lighter colors represent the VAD with 5\% false negatives. As the switching processing does not depend on the VAD, it is only shown for one case.
The error bars represent the standard deviation over all utterances. For the SNR improvement, the gray line and shaded area depict the mean SNR improvement of the nose pad microphone relative to the reference microphone and its standard deviation, since it has the highest input SNR.

The results for the SNR improvement on the left side of \cref{fig:ResultsEval} show that all beamformers yield an improvement compared to the nose pad microphone, where the adaptive and switching-adaptive beamformer perform similarly with an improvement of more than 10 dB compared to the best microphone for the oracle VAD, while the purely switching beamformer only yields an improvement of about 4-5 dB. In terms of SNR improvement, the performance of all algorithms is rather constant over the number of switches in the occlusion state and input SNR. 
For a VAD with 5\% false negatives, the performance of the adaptive and switching-adaptive beamformer clearly decrease compared to the oracle VAD. While at a low input SNR of 0 dB the two adaptive beamformers still outperform the purely switching beamformer, at high SNRs the performance drop due to an erroneous VAD is more severe and the performance of the adaptive and switching-adaptive beamformer decreases to the input SNR in the best microphone.
As already discussed in \cref{sec:Proc}, the results show the robustness of the purely switching beamformer against VAD errors, while also indicating the potential of adaptation when a good VAD is available.

In terms of own-voice distortions, it can be observed that the purely switching beamformer yields rather constant low distortions (around -10 dB) for all input SNRs and numbers of switches. The switching-adaptive beamformer with an oracle VAD (solid pink) constantly leads to even slightly lower distortions. The adaptive beamformer (blue) performs similarly to the switching-adaptive beamformer in terms of OVD for a low number of switches, while inducing more distortions if the occlusion state switches often (24 and 48 switches per utterance). Overall, the distortions for the two adaptive beamformers are slightly lower at a higher input SNR.
This result can be interpreted such that the purely adaptive beamformer is not capable of tracking the switches in the occlusion state, which is particularly noticeable if many switches occur at a high rate. There it seems beneficial to adapt two different sets of covariance matrices for the respective occlusion state to account for the fast dynamics in the occlusion pattern.

For an erroneous VAD, the observation for the OVD is similar to the SNR improvement: Overall, the performance decreases for the two adaptive beamformers, i.e., larger distortions are induced, where again this effect becomes more pronounced for high input SNRs. The above described trend of the switching-adaptive beamformer leading to lower distortions than the purely adaptive beamformer for highly dynamic occlusion patterns can also be observed here. 
The results for the OVDs further underline the robustness of the switching beamformer against VAD errors and the potential benefit of the adaptive beamformers if a good VAD is available.

\section{Conclusions}

In this paper, we investigated the influence of microphone occlusion and proposed different methods based on adaptive and switching beamformers, and a hybrid switching-adaptive beamformer. The proposed methods depend on an occlusion detection and/or a voice activity detection. 
The evaluation showed the advantages in terms of robustness for the purely switching beamformer, while also showing the potential benefits in terms of SNR improvement and own-voice distortions of adaptive beamforming if a good VAD is available. The advantage of the hybrid switching-adaptive beamformer could be demonstrated for fast dynamics in the occlusion pattern where less distortions were observed than for the adaptive beamformer, while performing similarly for relatively static occlusion patterns.
Since the switching-adaptive beamformer comes at the cost of higher computational complexity and memory consumption, in practical use cases, this trade-off should further be considered along with the occurring dynamics in the occlusion pattern.

\clearpage
\bibliographystyle{IEEEtran}
\bibliography{mybib.bib}

\end{document}